\documentclass[superscriptaddress,
  aip,
  pop,
  amsmath,amssymb,
  reprint,
]{revtex4-1}
\usepackage{graphicx,color}
\usepackage{amsmath}
\usepackage{natbib}
\usepackage{epsfig}
\begin{document}

\title{Note: Melting criterion for soft particle systems in two dimensions}

\author{Sergey Khrapak}
%\email{Sergey.Khrapak@dlr.de}
\affiliation{Institut f\"ur Materialphysik im Weltraum, Deutsches Zentrum f\"ur Luft- und Raumfahrt (DLR), 82234 We{\ss}ling, Germany}
\affiliation{Aix Marseille University, CNRS, Laboratoire PIIM, 13397 Marseille, France}
\affiliation{Joint Institute for High Temperatures, Russian Academy of Sciences, 125412 Moscow, Russia}

\date{\today}

\maketitle

According to the Berezinskii-Kosterlitz-Thouless-Halperin-Nelson-Young (BKTHNY) theory,~\cite{KosterlitzRMP2017} melting in two dimensions (2D) is a two-stage process. The crystal first melts by dislocation unbinding to an anisotropic hexatic fluid and then undergoes a continuous transition into isotropic fluid. The dislocation unbinding occurs when the Young's modulus reaches the universal value of $16\pi$,
\begin{equation}\label{KTHNY}
\frac{4\mu(\mu+\lambda)}{2\mu+\lambda}\frac{b^2}{k_{\rm B}T}= 16\pi,
\end{equation}
where $\mu$, $\lambda$ are the Lam{\' e} coefficients of the 2D solid, $b$ is the lattice constant, and $k_{\rm B}T$ is the thermal energy. The Lam{\' e} coefficients to be substituted in Eq.~(\ref{KTHNY}) should be evaluated taking into account (i) {\it thermal softening} and (ii) {\it renormalization} due to dislocation-induced softening of the crystal.~\cite{MorfPRL1979,ZanghelliniJPCM2005}   
Simplistic theoretical estimates using the elastic constants of an ideal crystalline lattice at $T=0$ yield melting temperatures overestimated by a factor between $\simeq 1.5$ and $\simeq 2$ for various 2D systems.~\cite{ZanghelliniJPCM2005,ThoulessJPC1978,PeetersPRA1987,KhrapakCPP2016}  

BKTHNY scenario has been confirmed experimentally, in particular for systems with dipole-like interactions.~\cite{ZahnPRL1999,GrunbergPRL2004,ZanghelliniJPCM2005} More recently, it has been reported that 2D melting  scenario depends critically on the potential softness.~\cite{KapferPRL2015} Only for sufficiently soft long-range interactions does melting proceed via the BKTHNY scenario. For steeper interactions the hard-disk melting scenario with first order hexatic-liquid transition holds.~\cite{BernardPRL2011,EngelPRE2013,ThorneyworkPRL2017}

The focus of this Note is on 2D soft particle systems. It is demonstrated that a melting criterion can be introduced, which states that the melting occurs when the ratio of the transverse sound velocity of an ideal crystalline lattice to the thermal velocity reaches a certain quasi-universal value. 

The Lam{\' e} coefficients of an ideal 2D lattice can be expressed in terms of the longitudinal ($C_{\rm L}$) and transverse ($C_{\rm T}$) sound velocities as $\mu = m\rho C_{\rm T}^2$ and $\lambda=m\rho(C_{\rm L}^2-2C_{\rm T}^2)$, where $m$ and $\rho$ are the particle mass and number density.~\cite{PeetersPRA1987,LL_Elasticity} Then the condition (\ref{KTHNY}) can be rewritten as 
\begin{equation}\label{Cond1}
2 \pi \sqrt{3} v_{\rm T}^2 = C_{\rm T}^2\left(1-C_{\rm T}^2/C_{\rm L}^2\right),
\end{equation}
where $v_{\rm T}=\sqrt{k_{\rm B}T/m}$ is the thermal velocity. For soft repulsive potentials, independently of space dimensionality,  the following strong inequality, $C_{\rm L}^2/ C_{\rm T}^2 \gg 1$, holds.~\cite{QCA_Relations,KhrapakSciRep2017,ElasticModuli} This implies that Eq.~(\ref{Cond1}) can be further simplified to 
\begin{equation} \label{Cond2}
C_{\rm T}/v_{\rm T} \simeq {\rm const}
\end{equation}
at melting. The value of the constant that follows from Eq.~(\ref{Cond1}) is $\simeq 3.30$. However, this does not take into account thermal and dislocation induced softening. A working hypothesis to be verified is that a simple renormalization of the constant in Eq.~(\ref{Cond2}) can account for these effects. In this case, Eq.~(\ref{Cond2}) would be identified as a simple 2D universal melting rule for soft particle systems. 

\begin{table}
\caption{\label{Tab1} Selected properties of 2D one-component plasma with logarithmic (OCP log), Coulomb (OCP $1/r$) interactions, and of the 2D system with the dipole-like interaction. Here $C_{\rm T}$ is the transverse sound velocity of an ideal triangular lattice, $v_{\rm T}$ is the thermal velocity, and $\Gamma_{\rm m}$ is the coupling parameter at melting.}
\begin{ruledtabular}
\begin{tabular}{crrrr}
System & $f(x)$ & $C_{\rm T}/v_{\rm T}$~\footnote{See e.g. Ref.~\onlinecite{Alastuey1981} for OCP log, Ref.~\onlinecite{PeetersPRA1987} for OCP $1/r$, and Ref.~\onlinecite{IPL3} for the dipole system.} & $\Gamma_{\rm m}$~\footnote{See Refs.~\onlinecite{CaillolJSP1982,Leeuw1982} for OCP log; Refs.~\onlinecite{GrimesPRL1979,GannPRB1979} for OCP $1/r$, and Refs.~\onlinecite{ZahnPRL1999,IPL3} for the dipole system.}  & $C_{\rm T}/v_{\rm T}|_{\Gamma_{\rm m}}$  \\ \hline
OCP log & $-\ln x$ & $\sqrt{\Gamma/8}$ & $\simeq 130 \div 140$
& $\simeq 4.0\div 4.2$   \\
OCP $1/r$ & $1/x$ &  $0.372\sqrt{\Gamma}$& $\simeq 120 \div 140$ & $\simeq 4.1 \div 4.4$ \\
Dipole & $1/x^3$ & $0.547\sqrt{\Gamma}$  & $\simeq 60 \div 70$ & $\simeq 4.2 \div 4.6$ \\
\end{tabular}
\end{ruledtabular}
\end{table}

Let us verify whether the ratio $C_{\rm T}/v_{\rm T}$ does assume a universal value at melting. We consider three exemplary  2D systems with soft long-ranged repulsive interactions: one-component plasmas with logarithmic potential (OCP log),~\cite{Alastuey1981,CaillolJSP1982,Leeuw1982} 2D electron system with Coulomb $\propto 1/r$ potential (OCP $1/r$),~\cite{GrimesPRL1979,GannPRB1979} , and dipole-like system with $\propto 1/r^3$ interaction.~\cite{ZahnPRL1999,GrunbergPRL2004,IPL3} The pair-wise interaction potential $\phi(r)$ can be written in a general form as 
\begin{displaymath}
\phi(r)/k_{\rm B}T = \Gamma f(r/a),
\end{displaymath} 
where $\Gamma$ is the coupling parameter and $a=1/\sqrt{\pi \rho}$ is the 2D Wigner-Seitz radius. The system is usually referred to as strongly coupled when $\Gamma \gg 1$. The fluid-solid phase transition is characterized by a system-dependent critical coupling parameter $\Gamma_{\rm m}$ (the subscript ``m'' refers to melting). All systems considered here form hexagonal lattices in the crystalline phase (more complicated interactions and lattices should be considered separately). 

The  discussed soft-particle systems have been extensively studied in the literature and some relevant information is summarized in Table~\ref{Tab1}. In particular, the last column lists the ratios  $C_{\rm T}/v_{\rm T}$ at melting. The values presented indicate that  as the potential steepness grows some weak increase of the ratio $C_{\rm T}/v_{\rm T}$ at melting is likely.  At the same time, all the values are scattered in a relatively narrow range, $4.3\pm 0.3$. This can justify using Eq.~(\ref{Cond2}) as an approximate one-phase criterion of melting of 2D crystals with soft long-ranged interactions.
 
\begin{figure}
\includegraphics[width=7cm]{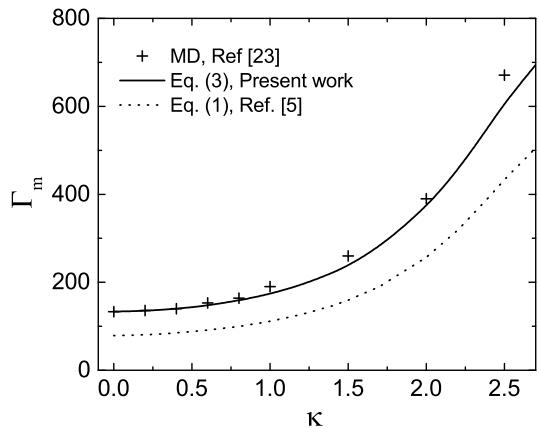}
\caption{Melting curve of a 2D Yukawa crystal in the ($\kappa$, $\Gamma$) plane. The solid curve corresponds to the condition $C_{\rm T}=4.3 v_{\rm T}$. The symbols correspond to the results of the numerical melting ``experiment''.~\cite{HartmannPRE2005} The dotted line corresponds to the solution of Eq.~(\ref{KTHNY}) with the asymptotic $T=0$ values of elastic constants.~\cite{PeetersPRA1987}}
\label{Fig1}
\end{figure}
 
As an example of the application of the proposed criterion, the melting curve of a 2D Yukawa crystal has been calculated. The Yukawa potential is characterized by $f(x)=\exp(-\kappa x)/x$, where $\kappa$ is the screening parameter (ratio of the mean interparticle separation $a$ to the screening length). This potential is used as a reasonable first approximation to describe actual interactions in colloidal suspensions and complex (dusty) plasmas.~\cite{FortovUFN,FortovPR,IvlevBook,ChaudhuriSM} In the latter case, the screening is normally weak,~\cite{NosenkoPRL2004,NosenkoPRL2009} $\kappa \lesssim 1$, which corresponds to the soft interaction limit.
Thermodynamics and dynamics of 2D Yukawa systems are well understood,~\cite{KalmanPRL2004,HartmannPRE2005,DonkoJPCM2008} simple practical expression for thermodynamic functions~\cite{KryuchkovJCP2017} and sound velocities~\cite{SemenovPoP2015,ElasticModuli} have been derived. In particular, the transverse sound velocity of an ideal Yukawa lattice can be expressed using the Madelung constant $M(\kappa)$ as~\cite{ElasticModuli}
\begin{displaymath}
C_{\rm T}^2=\frac{v_{\rm T}^2}{8}\left(\kappa^2\frac{\partial^2 M}{\partial \kappa^2}+\kappa\frac{\partial M}{\partial \kappa}-M\right),
\end{displaymath}   
where the product $M\Gamma$ defines the lattice energy per particle in units of temperature (reduced lattice sum). Using the condition (\ref{Cond2}) with an ``average'' ${\rm const} = 4.3$, the dependence $\Gamma_{\rm m}(\kappa)$ for the triangular lattice has been calculated. The resulting melting line (solid curve) appears in Fig.~\ref{Fig1}. The agreement with the numerical melting ``experiment''~\cite{HartmannPRE2005} is satisfactory in the weakly screened regime.  An early attempt to estimate the location of the melting curve by using Eq.~(\ref{KTHNY}) with the asymptotic $T=0$ values of elastic constants~\cite{PeetersPRA1987} is depicted by the dotted curve. A significant improvement is documented.               

To conclude, a simple criterion for melting of two-dimensional crystals with soft long-ranged interactions has been proposed. It states that the ratio of the transverse sound velocity of an ideal crystalline lattice to the thermal velocity is a quasi-universal number close to $4.3$ at melting. Application of this criteria allows estimating melting lines in a simple yet relatively accurate manner. Two-dimensional weakly screened Yukawa systems represent just one relevant example.   

Work at AMU was supported by the A*MIDEX project (Nr.~ANR-11-IDEX-0001-02). I thank Hubertus Thomas for reading the manuscript.

\bibliographystyle{aipnum4-1}
\bibliography{Melting2D_References}

\end{document}